\documentclass[12pt]{iopart}

\pdfoutput=1
\usepackage[backend=bibtex,style=numeric,sorting=none]{biblatex}
\DeclareFieldFormat{eprint:arxiv}{%
  Preprint available at \url{https://arxiv.org/abs/#1}%
}

\usepackage[pdftex]{graphicx}
\usepackage{amssymb}
\usepackage{enumitem}

\usepackage{algorithm, algorithmicx, algpseudocode}
\algnewcommand\INPUT{\item[\textbf{Input:}]}%
\algnewcommand\OUTPUT{\item[\textbf{Output:}]}%

\usepackage[linktocpage,breaklinks]{hyperref}
\usepackage{url}
\usepackage{booktabs,multirow}
\usepackage[all]{hypcap}
\usepackage{listings}
\usepackage[dvipsnames]{xcolor}
\graphicspath{{./}}
\usepackage{etoolbox}
\usepackage{siunitx}

\usepackage{hyperref}
\usepackage{orcidlink}

\hypersetup{colorlinks=true,
            citecolor=RoyalBlue,
            linkcolor=RoyalBlue,
            urlcolor=RoyalBlue}

\usepackage{xspace}
\usepackage{array}
\usepackage{dcolumn}
\newcolumntype{d}[1]{D{.}{.}{#1}}

\makeatletter
\newcommand{\mainmatter}{%
  \setcounter{footnote}{0}%
  \patchcmd{\@makefntext}{\fnsymbol}{\arabic}{}{}%
  \patchcmd{\@thefnmark}{\fnsymbol}{\arabic}{}{}%
  \def\@makefnmark{\textsuperscript{\arabic{footnote}}}%
  \long\def\@makefntext##1{\noindent\hbox{\textsuperscript{\arabic{footnote}}}\,\,##1}%
}
\makeatother

\newcommand{\orcid}[1]{\href{https://orcid.org/#1}{\includegraphics[height=\fontcharht\font`\B]{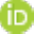}}}

\newcommand{\uidaho}{Department of Physics, University of Idaho, Moscow, ID 83843, USA}
\newcommand{\wvu}{Department of Physics and Astronomy, West Virginia University, Morgantown, WV 26506, USA}
\newcommand{\cgwc}{Center for Gravitational Waves and Cosmology, West Virginia University, Chestnut Ridge Research Building, Morgantown, WV 26505}
\newcommand{\cguwm}{Center for Gravitation, Cosmology and Astrophysics, Department of Physics, University of Wisconsin-Milwaukee, Milwaukee, WI 53211, USA}

\newcommand{\nrpy}{{\texttt{NRPy}}\xspace}
\newcommand{\nell}{{\texttt{NRPyElliptic}}\xspace}
\newcommand{\nellg}{{\texttt{NRPyEllipticGPU}}\xspace}

\newcommand{\etk}{\texttt{Einstein Toolkit}\xspace}
\newcommand{\bhah}{{\texttt{BlackHoles@Home}}\xspace}

\newcommand{\cuda}{{\texttt{CUDA}}\xspace}
\newcommand{\hip}{{\texttt{HIP}}\xspace}
\newcommand{\sycl}{{\texttt{SYCL}}\xspace}
\newcommand{\raja}{{\texttt{RAJA}}\xspace}
\newcommand{\kokkos}{{\texttt{Kokkos}}\xspace}
\newcommand{\openmp}{{\texttt{OpenMP}}\xspace}
\newcommand{\openacc}{{\texttt{OpenACC}}\xspace}
\newcommand{\simd}{{\texttt{SIMD}}\xspace}
\newcommand{\simt}{{\texttt{SIMT}}\xspace}
\newcommand{\SymPy}{\texttt{SymPy}\xspace}

\newcommand{\pone}{{Paper I}\xspace}

\newcommand{\host}{\textit{host}\xspace}

\newcommand{\device}{\textit{device}\xspace}

\newcommand{\grid}{\texttt{Grid}\xspace}
\newcommand{\block}{\texttt{Block}\xspace}
\newcommand{\sm}{\texttt{sm}\xspace}
\newcommand{\stream}{\texttt{stream}\xspace}
\newcommand{\nstreams}{\texttt{nstreams}\xspace}

\newcommand{\SinhSymTP}{{\texttt{SinhSymTP}}\xspace}

\newcommand{\eqref}[1]{Eq.\,(\ref{#1})}
\renewcommand{\algref}[1]{Alg.\,(\ref{#1})}

\def\eadnew#1#2{\address{#2 Corresponding Author: \mailto{#1}}}

\newcommand{\GridRes}[3]{\mbox{\ensuremath{#1{\times}#2{\times}#3}}\xspace}

\definecolor{keywordcolor}{rgb}{0.5,0,0.5}
\definecolor{commentcolor}{rgb}{0.2,0.6,0.2}
\definecolor{stringcolor}{rgb}{0.6,0.1,0.1}

\lstdefinelanguage{CUDA}{
    language=C++, 
    morekeywords={__global__, __device__, __shared__, __host__, <<<, >>>}, 
    sensitive=true, 
    moredelim=[s][\color{keywordcolor}]{<<<}{>>>}, 
}

\begin{document}
\sloppy 

\mainmatter

\date{\today}
\title[S. Tootle et al]
{Accelerating Numerical Relativity with Code Generation: CUDA-enabled Hyperbolic Relaxation}
\author{
Samuel~D.~Tootle$^{1\,,*}$~\orcid{0000-0001-9781-0496},
Leonardo~R.~Werneck$^1$~\orcid{0000-0002-4541-8553},
Thiago~Assump\c{c}\~{a}o$^{2\,,3\,,4}$~\orcid{0000-0002-3419-892X},
Terrence~Pierre~Jacques$^{1\,,3\,,4}$~\orcid{0000-0002-8993-0567},
Zachariah~B.~Etienne$^{1\,,3\,,4}$~\orcid{0000-0002-6838-9185}
}

\address{$^1$ \uidaho}
\address{$^2$ \cguwm}
\address{$^3$ \wvu}
\address{$^4$ \cgwc}
\eadnew{sdtootle@uidaho.edu}{$^{*}$}

\begin{abstract}
  Next-generation gravitational wave detectors such as Cosmic Explorer,
  the Einstein Telescope, and LISA, demand highly accurate and extensive
  gravitational wave (GW) catalogs to faithfully extract physical
  parameters from observed signals. However, numerical relativity (NR)
  faces significant challenges in generating these catalogs at the
  required scale and accuracy on modern computers, as NR codes do not
  fully exploit modern GPU capabilities. In response, we extend \nrpy, a
  Python-based NR code-generation framework, to develop \nellg---a
  CUDA-optimized elliptic solver tailored for the binary black hole (BBH)
  initial data problem. \nellg is the first GPU-enabled elliptic solver
  in the NR community, supporting a variety of coordinate systems and
  demonstrating substantial performance improvements on both
  consumer-grade and HPC-grade GPUs. We show that, when compared to a
  high-end CPU, \nellg achieves on a high-end GPU up to a sixteenfold
  speedup in single precision while increasing double-precision
  performance by a factor of 2--4. This performance boost leverages the
  GPU's superior parallelism and memory bandwidth to achieve a
  compute-bound application and enhancing the overall simulation efficiency.
  As \nellg shares the core infrastructure common to NR codes, this work
  serves as a practical guide for developing full, \cuda-optimized NR
  codes.
\end{abstract}
\submitto{\CQG}
\maketitle

\section{Introduction}
\label{sec:intro}

Numerical relativity (NR) plays a crucial role in the prediction,
detection, and interpretation of signals observed in multi-messenger
astronomy. The LIGO/Virgo/KAGRA Collaboration continues to detect compact
object binary mergers, with the majority of signals consistent with
binary black hole (BBH)
mergers~\cite{LIGOScientific:2021usb,KAGRA:2021vkt}. The accuracy of
gravitational wave (GW) models, as well as the template banks used by
observatories to determine whether a GW event has been detected, heavily
depends on the precision of NR
simulations~\cite{Gayathri:2020coq,Lange:2017wki,LIGOScientific:2016kms}.

NR simulations of GW sources form the foundation for extracting
parameters from observed GW signals, however, they come with a
significant computational cost.  This is reflected in recent
high-performance computing (HPC) usage statistics which show that NR and
GW physics are ranked as the third-highest consumer of HPC resources per
simulation in recent years~\cite{web:access_metrics}. Looking ahead, the
demand for computational resources in NR is expected to increase
significantly as third-generation (3G) GW detectors, such as LISA, Cosmic
Explorer, and the Einstein Telescope, become operational. These
instruments will necessitate a tenfold increase in simulation
accuracy~\cite{Purrer:2019jcp,Ferguson:2020xnm} and a significant
expansion of simulation catalogs to explore new GW sources, including
eccentric and high-mass-ratio compact binaries. Achieving these goals
with current NR codes would entail an impractically high computational
cost.

NR groups developing next--generation frameworks to meet the demands of
3G detectors face considerable challenges as advances in artificial
intelligence and graphical processing unit (GPU) technologies are rapidly
transforming the computing landscape. Modern HPC systems increasingly
adopt heterogeneous architectures, where GPUs perform the majority of the
computational workload rather than traditional central processing units
(CPUs). Thus, there is an urgent need for NR frameworks capable of
efficiently utilizing these architectures.

This challenge extends beyond large-scale HPC environments to
consumer-grade hardware, where dedicated GPUs now offer remarkable
computational performance compared to native CPUs. This is particularly
relevant to our group's proposed \bhah volunteer computing project, which
aims to harness consumer-grade hardware for generating 3G-ready BBH GW
catalogs with full NR~\cite{Ruchlin:2017com, Etienne2024}.

BBH systems form the foundation of compact binary GW data analysis.
Simulating such a system involves two basic steps: first, solving a set
of elliptic partial differential equations (PDEs) to compute the initial
data solution; and second, evolving this data forward in time by solving
a system of hyperbolic PDEs (see
e.g.,~\cite{BaumgarteShapiro_2010,Gourgoulhon:2007ue} for comprehensive
reviews). Both calculations are computationally intensive, with the
computational cost increasing for higher binary mass ratio and the
component spins of the BHs.

Reducing the cost of 3G-ready NR BBH simulation campaigns on modern
computing resources requires a deep understanding of both CPU and GPU
architectures, their strengths and limitations. Typically, existing
CPU-centric algorithms must be rewritten to fully utilize the latest
CPU/GPU heterogeneous architectures.
This presents a significant computational challenge in terms of
application portability, with the goal of enabling developers to write an
implementation once and have it perform efficiently across a broad
spectrum of computing technologies. Ref.~\cite{davis2024} provides a
recent evaluation that compares and contrasts native codes (i.e., those
developed using AMD's \hip or NVIDIA's \cuda) with several programming
models aimed at providing application portability, including
\openmp~\cite{openmp18}, \openacc~\cite{oacc14}, \sycl~\cite{sycl23},
\raja~\cite{raja19}, and \kokkos~\cite{kokkos22}.
To summarize, applications that effectively leverage these programming
models demonstrate promising performance across a wide range of HPC
resources. Notably, \kokkos and \sycl sometimes achieve superior
performance to native applications, but typically underperform.

In an effort to effectively leverage the latest HPC resources, the NR
community continues to develop GPU-accelerated applications, with a
primary focus on the time evolution of dynamical spacetimes. For general
relativistic magnetohydrodynamic (GRMHD) simulations of isolated objects,
known frameworks include \texttt{AsterX}~\cite{Kalinani2024} and
\texttt{GRaM-X}~\cite{Shankar:2022ful}, both of which use the new
\etk~\cite{etk_web} driver \texttt{CarpetX}, built on the
\texttt{AMReX}~\cite{AMReX_JOSS} programming framework. In contrast,
\kokkos has been employed to rewrite \texttt{Athena++} into more portable
implementations, including the multi-physics framework
\texttt{Parthenon}~\cite{grete2020k, Grete2023}, as well as
\texttt{AthenaK}, which supports simulations of BBHs~\cite{Zhu2024} and
GRMHD studies of binary and isolated neutron stars~\cite{Fields2024}. In
addition, NR frameworks natively ported using \cuda include
\texttt{SpEC}~\cite{Lewis:2018qqc}, optimized for simulating isolated
BHs, and \texttt{Dendro-GR}~\cite{Fernando:2022zbc}, designed for BBH
simulations.

An alternative to relying solely on portable programming libraries is
code generation, which has become a cornerstone in both industry and NR
applications~\cite{Ruchlin:2017com, Palenzuela:2018sly, Fernando:2022zbc,
Peterson2023a}. In this work, we describe the initial steps toward
extending \nrpy, a Python-based code-generation framework tailored for
NR~\cite{Ruchlin:2017com,nrpy_web}, to enable the creation of highly
optimized, GPU-accelerated NR applications for both consumer-grade and
HPC-grade GPUs. Specifically, we focus on adapting key components to
generate \nellg, a \cuda-optimized version of \nell, which leverages a
hybrid CPU+GPU approach to maximize computational efficiency while
retaining the adaptability of \nell to support an array of coordinate
geometries and compatibility with \nell's import thorns for the Einstein
Toolkit.

In a previous work~\cite{Assumpcao:2021fhq} (hereafter \pone), our group
introduced \nell, an extensible elliptic solver specifically designed for
NR problems and generated using \nrpy. In \pone, the capabilities of
\nell were demonstrated by solving the BBH initial data problem, a
critical step in simulating GW sources. Leveraging the hyperbolic
relaxation method~\cite{Ruter:2017iph}, \nell solves nonlinear elliptic
PDEs using a single grid with compactified curvilinear coordinates that
exploit near-symmetries of the underlying physical system. \nell also
implements an adaptive relaxation wavespeed that accelerates relaxation
while maintaining the Courant-Friedrichs-Lewy (CFL) stability condition.
To further enhance performance, \nell incorporates \openmp
parallelization and advanced \simd instruction generation.

A significant advantage of the hyperbolic relaxation approach is that it
leverages the same numerical methods and code-generation infrastructure
used by NR evolution codes focused on the time evolution of dynamical
spacetimes. Thus, the impact of this paper is twofold: it introduces the
first GPU-enabled initial data solver for BBH ID and serves as the
foundation for \cuda-enabled NR evolution codes generated automatically.
One such application is \bhah, which specializes in modeling dynamical
spacetimes for isolated and binary compact objects.

To this end, this study rigorously evaluates the efficiency of \nell and
\nellg by analyzing their performance on both consumer-grade and
HPC-grade hardware. We also demonstrate roundoff-level agreement between
\nell and \nellg at double precision. While previous GPU-based NR
analyses have largely focused on HPC-grade hardware, even for single-node
performance~\cite{Fernando:2022zbc, Shankar:2022ful, Lewis:2018qqc},%
\footnote{Notable exceptions include Refs.~\cite{Zhu2024}, which
considers only single-core performance, and~\cite{Fields2024}.}
our work addresses this limitation by exploring GPU optimizations for NR
applications across a broader range of hardware. This approach emphasizes
both accessibility and scalability, bridging the gap between
consumer-grade and high-performance systems.

The remainder of this paper is organized as follows:
Sec.~\ref{sec:basic_equations} provides a brief overview of the elliptic
system used to model the two-puncture problem and the hyperbolic
relaxation method employed to solve it. In Sec.~\ref{ssec:gpu_overview},
we describe the relevant features of the \cuda programming model,
followed by a detailed overview of the numerical implementation and
design decisions to extend \nrpy for generating optimized \cuda code in
Sec.~\ref{ssec:gpu_adapt}. Section~\ref{sec:results} presents our
results, including agreement between \nellg and \nell to roundoff
precision, additional enhancements to achieve the final benchmarks, and
rigorous analysis of hardware-imposed limitations. Finally,
Sec.~\ref{sec:conclusions} discusses the lessons learned, outlines a path
toward optimized \cuda applications for dynamical spacetimes, and
provides plans for future work. A glossary of acronyms can be found in
Tab.~\ref{tab:acronyms}.

\section{Basic equations}
\label{sec:basic_equations}

Since the basic equations were introduced in detail in \pone, we provide
only a brief overview here. Using \nell and \nellg, we solve the system
of equations to construct conformally flat BBH initial data. Our starting
point is the $3+1$ Arnowitt-Deser-Misner (ADM) decomposition of the
spacetime manifold~\cite{Arnowitt:1959ah, Arnowitt:1962hi}:
\begin{equation}
  ds^2 = -\alpha^2 dt^2 + \gamma_{ij}(dx^i + \beta^i dt)(dx^j + \beta^j dt)\,,
\end{equation}
where $\gamma_{ij}$ is the physical spatial metric, $\alpha$ is the lapse
function, and $\beta^{i}$ is the shift vector. The conformal
transverse-traceless (CTT) decomposition (see, e.g.,~\cite{Cook:2000vr})
reformulates the physical metric using the conformal decomposition:
\begin{equation}
  \gamma_{ij} = \psi^4\tilde{\gamma}_{ij} \,,
\end{equation}
where $\psi$ is a positive scalar function (the conformal factor) and
$\tilde{\gamma}_{ij}$ is the conformal metric. Under the assumption of
asymptotic flatness, $\tilde{\gamma}_{ij}$ is set to the flat,
three-dimensional metric $\hat{\gamma}_{ij}$. For
Brandt-Br\"ugmann~\cite{Brandt:1997tf} puncture initial data considered
here, this results in an analytic solution to the momentum constraint
equations, leaving only the Hamiltonian constraint to solve. Because the
conformal factor is singular, it is decomposed into singular and
non-singular components:
\begin{equation}
  \psi = \psi_{\rm singular} + u \,.
\end{equation}
Substituting this into the Hamiltonian constraint, the equation for the
non-singular function $u$ becomes:
\begin{eqnarray}
  \label{eqn:hamiltonian_constraint_v3}
  \hat{\nabla}^2 u
  + \frac{1}{8} \tilde{A}_{ij} \tilde{A}^{ij} (\psi_{\rm singular} + u)^{-7} = 0 \,,
\end{eqnarray}
where $\hat{\nabla}$ is the covariant derivative compatible with the flat
background metric $\hat{\gamma}_{ij}$ and $\tilde{A}_{ij}$ is the
conformal trace-free extrinsic curvature. We solve this equation using
the hyperbolic relaxation method, which
recasts~\eqref{eqn:hamiltonian_constraint_v3} as a system of coupled
first-order (in time) PDEs:
\begin{eqnarray}
  \partial_{t}u &=& v - \eta u \,, \nonumber \\
  \partial_{t}v &=& c^{2}\left[\tilde{\nabla}^{2}u
    + \frac{1}{8}\tilde{A}_{ij}\tilde{A}^{ij}(\psi_{\rm singular}
    + u)^{-7} \right]
  \,, \label{eq:two_punctures_rfm_system}
\end{eqnarray}
where $\eta$ is a damping factor and $c$ is the wave speed. In the
steady-state regime, where $\partial_{t}u = \partial_{t}v = 0$, the
solution of Eqs.~\ref{eq:two_punctures_rfm_system} coincides with that of
the original system.

Formulated as a hyperbolic PDE, this approach to solving elliptic PDEs
completely avoids the need to recast the PDEs in matrix form or to
linearize them. Further, being hyperbolic, it adopts the same
infrastructure as a traditional NR evolution code, making implementation
even easier. However, this convenience usually comes at a cost:
hyperbolic relaxation methods are relatively inefficient for solving
elliptic PDEs, requiring that constant-speed relaxation waves cross the
numerical grid many times.

As described in \pone, we address this inefficiency by adjusting the wave
speed $c$ in Eqs.~\ref{eq:two_punctures_rfm_system} to be proportional to
the local grid spacing and by utilizing NRPy's \SinhSymTP coordinate
system.  These tools provide a nonuniform grid that is denser near the
punctures and grows exponentially toward the outer boundary while
ensuring that the CFL constraint is satisfied throughout the relaxation.
Cumulatively, these modifications significantly accelerate relaxation
waves toward the outer boundary thereby drastically reducing the
relaxation time, bringing this method to the level of state-of-the-art
elliptic solvers.

Having established the foundational equations and the numerical method
used to solve them, we next outline the core requirements for adapting NR
codes like \nell to GPUs. Recognizing that the reader, likely a numerical
relativist, may not be familiar with the intricacies of \cuda
programming, we begin by providing an overview of the \cuda programming
model.

\section{Overview of the \cuda programming model}
\label{ssec:gpu_overview}

To leverage \cuda-enabled GPUs effectively for NR applications, it is
essential to understand the fundamental components and principles of the
\cuda programming model. To do so, we will adopt the standard \cuda
coding naming conventions throughout this text such that references to
\device (\host) code refer to code executed by the GPU (CPU). Developed
by NVIDIA, \cuda is a parallel computing platform that enables the use of
NVIDIA GPUs for general-purpose computations, significantly accelerating
tasks that benefit from parallel processing.

At the core of \cuda are \textit{Streaming Multiprocessors} (SMs), which
are analogous to CPU cores, but designed to handle thousands of
lightweight threads concurrently. While a typical CPU may have a handful
of powerful cores optimized for sequential or vectorized processes as well as complex logic,
an SM contains multiple smaller cores optimized for high-throughput
parallelism. This architecture is supported by a hierarchical memory
system, similar to the multi-level cache hierarchy in CPUs, designed to
optimize data access and hide latency.

Each generation and class of \cuda-enabled device has an associated
compute capability which corresponds to intrinsic features and
specifications that are available in NVIDIA's \textit{C Programming
Guide}~\cite{nvidia_cpg}. A critical insight into a GPU's capabilities
(see section 5.4.1 of~\cite{nvidia_cpg}) is the number of arithmetic
instructions (for a given precision) that can be performed per clock
cycle per SM. Understanding these specifications is important to
determine the maximum theoretical performance of an application on a
given GPU, a point that will be revisited in Sec.~\ref{sec:results}.

A fundamental concept in \cuda is the \textit{Single Instruction,
Multiple Threads} (\simt) model, which shares similarities with the
\textit{Single Instruction, Multiple Data} (\simd) model used in CPUs,
but offers greater flexibility. For \cuda-enabled devices, threads are
grouped into units called \textit{warps} (typically 32 threads) that must
execute the same instruction. However, each thread in a warp can follow
its own control flow which can result in threads within a warp taking a
different execution path (a situation known as thread divergence).  In
this case, performance can degrade substantially. Therefore, minimizing
thread divergence is crucial for optimizing performance, analogous to
avoiding branch mis-predictions and ensuring uniform execution in \simd
operations on CPUs.

Within an SM, threads are organized into \textit{blocks}, which
determines the number of active threads per SM, the number of registers
available per thread, and the tiling strategy for looping over grid data.
Blocks are further organized into a \textit{grid}, which represents the
3D domain of the problem and is used to distribute the workload across
the GPU. To maximize the computational performance of each thread, it is
essential to ensure each thread has the maximum available cache
resources.  For the GPUs used here, this corresponds to 255 32-bit
registers per thread with a block size not to exceed 256 threads.

There are four key types of memory in \cuda pertinent to this work:
\textit{global}, \textit{shared}, \textit{registers}, and
\textit{constant}. \textbf{Global memory} offers larger storage capacity
for GPUs, and is analogous to the main system memory (RAM) accessible by
all CPU cores, though with significantly higher bandwidth and
significantly lower capacity. Much like RAM, accessing global memory is
typically much slower than accessing caches or registers. Efficient use
of global memory often requires organizing data so that consecutive
threads access contiguous memory locations, maximizing data throughput
and minimizing latency, similar to optimizing memory access patterns in
CPU applications to leverage cache lines effectively.

\textbf{Shared memory} is located at the SM level and is accessible by
all threads within the same block. It is significantly faster than global
memory and is ideal for storing data that multiple threads need to access
frequently, thereby reducing the need to fetch data from slower global
memory repeatedly. This is analogous to how CPU cores share caches to
speed up data access among threads, however, unlike CPUs shared memory is
explicitly managed by the programmer.

Each thread has its own set of \textbf{registers}, providing the fastest
access to data. This is analogous to CPU registers, which store
frequently accessed variables for quick computation. However, similar to
CPUs, the number of registers is limited, so efficient usage is crucial
to avoid spilling to slower memory and incurring performance penalties.

\textbf{Constant memory} is a specialized read-only memory area cached on
the GPU, with a small cache size of $\sim 64{\rm KB}$. Constant memory is
optimized for scenarios where all threads read the same data
simultaneously, much like broadcast instructions in CPU \simd operations,
with latency comparable to an L1 cache. While the allocation of
constant memory must be known at compile time, the stored data can be modified
in \host code at runtime. The \textit{constness} comes from being
immutable in \device code. Additionally, constant data can be implicitly
stored in the constant cache by the compiler if the compiler is able to
determine that a variable is constant at compile time and will fit into
the cache limits.

Within the \cuda application programming interface (API), execution on the GPU is achieved using
\textit{Global} kernels. Global kernels are central to \cuda programming
as they are the mechanism for the \host to launch code on the \device and
for defining a parallelization strategy.  As such, they must be declared
and launched in a specific manner:
\begin{lstlisting}[language=CUDA]
// Kernel function declaration (e.g., in a header file)
__global__ global_kernel(MyDataType data);

// Kernel launch by the host (e.g., in the main function)
global_kernel<<<Grid, Block, sm, stream>>>(data);
\end{lstlisting}
where, in reverse order, \stream is an optional argument that specifies
the kernel to execute using a specific \cuda stream; \sm is the optional
bytesize of shared memory arrays which allows for dynamic specification;
\block is a \device organizational structure that represents the number
of threads per block; and \grid specifies the number of logical
\texttt{Blocks} required to complete the calculation. Organizing threads
and blocks effectively is essential for maximizing parallel efficiency,
much like optimizing thread distribution and workload balance across CPU
cores to prevent bottlenecks and ensure efficient utilization.

It is important to note that execution of global kernels by the \host are
non-blocking. Specifically, after the kernels are launched by
the \host, the \host will continue executing the next instructions until
an explicit synchronization is encountered.  The execution of additional
global kernels by the \host will then be tasked to the GPU scheduler and
executed in the order they are tasked.

Data transfer between the \host and \device occurs over the PCI-Express
bus, introducing latency and bandwidth limitations. To mitigate these
issues, \cuda provides features such as \textit{pinned (page-locked)
memory} and asynchronous data transfers. Pinned memory ensures faster
data transfers by preventing the operating system from moving the memory
to slower storage, such as disk or swap space---a process called
``paging.'' By locking the memory in RAM, it guarantees that the data
remains immediately accessible for transfers. This approach is similar to
reserving specific memory regions in CPU applications to ensure
consistent and rapid access.

\cuda also supports \textit{streams}, which is an additional layer of
parallelization that allows multiple kernels or memory operations to be
executed concurrently. By assigning independent tasks to separate
streams, computation and communication operations can be executed
asynchronously, improving overall performance by allowing the CPU and GPU
to work concurrently without waiting for data transfers to complete.

Finally, \cuda offers \textit{intrinsics}, which are specialized
functions that provide optimized performance for specific mathematical
operations~\cite[see Sec.~13.2 in][]{nvidia_cpg}. These intrinsics allow
for more efficient computations by leveraging hardware-specific
instructions, enabling higher performance without relying solely on
compiler optimizations. This is akin to using \simd intrinsics functions
for CPU code that map directly to specific assembly instructions,
allowing one to exploit processor-specific features for performance
gains. Incorporating CUDA intrinsics can lead to significant speedups in
compute-intensive parts of NR applications, given NR calculations often
involve complex mathematical operations that can be challenging for
compilers to effectively optimize.

\subsection{Key Considerations for Numerical Relativity Applications}

When adapting NR codes to run on \cuda-enabled GPUs, several important
factors must be addressed to achieve optimal performance, drawing
parallels to optimization strategies employed on CPU architectures:
\begin{itemize}
    \item \textbf{Memory Management}: NR simulations often involve large
    datasets and complex data structures. Efficiently organizing data to
    take advantage of shared and constant memory can significantly reduce
    access times and improve performance, much like optimizing data
    layout in CPU caches to enhance cache locality and minimize cache
    misses.
    \item \textbf{Parallelization Strategy}: The inherent parallelism in
    NR problems must be mapped effectively to the GPU's architecture.
    This involves decomposing the computational domain and ensuring that
    the workload is evenly distributed across threads to prevent
    bottlenecks, analogous to distributing tasks evenly across CPU cores
    to maximize parallel efficiency and avoid core idle times. Tasks that
    are predominantly serial should be avoided, as they can lead to
    significant performance degradation.
    \item \textbf{Minimizing Divergence}: Conditional operations in NR
    algorithms can lead to thread divergence within warps, thereby
    reducing performance. Designing kernels to minimize these divergences
    ensures that all threads within a warp execute instructions
    efficiently, similar to minimizing branch instructions and ensuring
    consistent execution paths in CPU \simd operations to prevent
    pipeline stalls.
    \item \textbf{Data Transfer Optimization}: Reducing the frequency and
    volume of data transfers between the host and device is crucial.
    Techniques such as overlapping computation with data transfers and
    utilizing pinned memory can help mitigate the impact of PCI-Express
    latency.
    \item \textbf{Scalability}: Ensuring that the code scales well with
    increasing problem sizes and fully utilizes the computational power
    of modern GPUs, including both consumer-grade and HPC-grade hardware,
    is vital for future-proofing NR simulations. This is comparable to
    designing CPU applications that scale efficiently with the number of
    cores and leverage advanced CPU features to maintain performance as
    hardware evolves.
\end{itemize}

Addressing these considerations enables NR applications to fully exploit
the computational capabilities of \cuda-enabled GPUs.  In the following
section, we will discuss how these are tackled in the extension of
\nrpy to generate high-performance NR applications.

\section{Adapting \nrpy for \cuda code generation}
\label{ssec:gpu_adapt}

Building on these foundational principles of the \cuda programming model
and its application to NR, we now explore the extension of \nrpy to
generate \cuda-optimized NR codes. Our primary focus is on \nellg, which
is the first application to fully leverage these new features. We begin
by providing an \emph{Algorithmic Overview} of \nellg, including its
hybrid CPU+GPU design and the structural inheritance from \nrpy's native
infrastructure, \bhah. Next, we discuss how \nellg addresses each of the
five key considerations when adapting NR applications to GPUs, as
discussed in the previous section. Finally, we present the complete
\nellg algorithm (Alg.~\ref{alg:hybriddriver}), which we reference
throughout this section.

Adapting \nrpy for generating \cuda-enabled applications involves a
reimagination of the code generation process such that the key
abstractions such as the grid and mathematical calculations are preserved
while the underlying implementation is optimized for GPU execution by
leveraging the unique features of the \cuda programming model. Although
the underlying extensions are generic and reusable, we use \nellg as an
illustrative example. \nrpy's \bhah infrastructure, used by \nellg/\nell
and detailed in \cite{Ruchlin:2017com,Assumpcao:2021fhq}, is particularly
well-suited for GPU acceleration for three main reasons:
\begin{enumerate}
  \item \textbf{Memory-Efficient, Multipatch Design:} \bhah's native
  support for curvilinear coordinates minimizes memory usage, making even
  consumer-grade GPUs with only 8--14GB of memory sufficient for
  nontrivial NR simulations.
  \item \textbf{Flattened Grid Arrays:} By storing all grid functions in
  a single flat array, \nrpy reduces pointer overhead and balances GPU
  register usage more effectively than if each grid function were
  separately allocated.
  \item \textbf{Large, Compute-Intensive Kernels:} \nrpy naturally
  generates sizable kernels involving finite-difference stencils, which
  GPUs handle efficiently thanks to their high memory bandwidth and \simt
  execution model.
\end{enumerate}

In addition, we extend \nrpy to leverage a hybrid approach to allow the
\host to handle small, sequential tasks while offloading compute-heavy
sections to the \device. In \nellg, this prevents performance degradation
from excessive \textit{host--device} data transfers and optimizes the
overall runtime. Specifically, the main GPU-intensive tasks in
\algref{alg:hybriddriver} include computing right-hand sides ({\bf RHS}),
Hamiltonian constraint residuals ({\bf H}), Runge-Kutta ({\bf RK})
stages\footnote{%
This is the calculation of $k_s$ for a given Runge-Kutta implementation,
i.e. after a RHS evaluation.}, %
and boundary conditions ({\bf BC}). Tasks that are inherently sequential
or require minimal computation, such as identifying boundary points
between grids and writing checkpoints, remain on the CPU side. As noted
in lines 8--9 and steps (a)--(c) of \algref{alg:hybriddriver}, this
approach ensures that if GPU scheduling and synchronization overhead
outweigh the benefit of parallelizing a small task, the task will be done
on \host and only the data needed will be transferred to the \device.

\subsection{Memory Management}

Effective memory management is crucial when porting NR codes to GPUs.
Since NR problems often involve large arrays of data, \nrpy's decision to
flatten grid functions into a single array reduces the overhead of
pointer dereferencing and helps maintain ample registers for arithmetic
operations. Beyond this foundational step, two additional strategies help
manage memory efficiently:

\begin{itemize}
  \item \textbf{Measuring Memory Requirements Early:} Since the average
  consumer GPUs today offer as little as 8\,GB of memory, it's essential
  to determine if the problem size fits within the available GPU memory.
  In \nrpy applications, the total memory footprint is allocated up front
  at runtime, thereby providing quick feedback on whether a chosen
  problem size fits on the GPU. This is particularly important as this
  minimizes memory (de)allocation overhead and because high-end consumer
  cards may have up to ${\sim}14$\,GB of RAM, while HPC-grade GPUs can
  exceed 40\,GB. Consequently, the memory limit remains a major concern
  for large NR applications.
  \item \textbf{Use of Constant Memory:} \cuda constant memory is
  leveraged for read-only, frequently accessed data. This is especially
  beneficial for small arrays of numerical parameters or precomputed
  constants (e.g., Runge-Kutta coefficients), which the compiler can
  store in a fast cache shared by all threads.
\end{itemize}

In \nrpy, we leverage constant memory in two ways. First, we explicitly
store constant parameters as (e.g. number of grid points, grid spacing,
dt, relaxation wavespeed) which are copied as needed to the \device prior
to launching a GPU kernel. In practice this storage ends up being arrays
of length \nstreams.  Second, we leverage the compiler's ability
to implicitly store numerical constants (e.g. Runge-Kutta coefficients)
in constant memory. This is achieved by using \SymPy and \nrpy's advanced
CSE algorithms to aggressively identify numerical constants and move them
to \texttt{const} (or \texttt{static constexpr} for C++ applications)
variable definitions. The added benefit is that expensive instructions to
compute rational constants can be moved to compile time and efficiently
accessed by all threads.

\subsection{Parallelization Strategy}
Under the \cuda paradigm, parallelization involves launching one or more
kernels over a grid of thread blocks, each block containing multiple
threads. \nrpy translates standard CPU loops into global \cuda kernels by
mapping loop indices to thread and block indices. This seamless
translation is facilitated by the flattened array representation, which
simplifies kernel logic.

By default, \nrpy uses a block size of $(32,\,\mathbf{NGHOSTS},\,1)$,
where 32 is the typical warp size and \mbox{$\mathbf{NGHOSTS}$} is the
radius of the finite-difference stencil. Although this choice is not
always optimal for every kernel, it balances performance across a variety
of possible finite-difference orders (2--12). Tests with
profiler-recommended block sizes (e.g., using \textit{NVIDIA Nsight
Compute}) showed marginal speedups, emphasizing that current compute
limitations often arise from hardware constraints and double-precision
demands.

Although \cuda supports dynamic allocation of on-chip shared memory, we
do not currently rely on it for \nellg. Tests that included shared memory
strategies did not significantly improve performance for the
compute-heavy kernels, which are currently limited by double-precision
hardware throughput rather than memory bandwidth. Specifically, using
NVCC, we observed that using shared memory optimized kernels would reduce
the number of generated instructions without a measurable speedup for the
compute-bound parts of the code. It also introduced significant code
complexity in \nrpy for generating such kernels. However, shared memory
may prove more beneficial for future multi-patch evolution codes (e.g.,
BSSN) or other multi-kernel workflows.  We plan to revisit shared memory
strategies when exploring those more complex applications.

\subsection{Minimizing Divergence}

In \cuda's \simt execution model, threads within a warp execute
instructions in lockstep. If threads within the same warp follow
different control flows (branching), performance degrades due to warp
divergence. \nrpy uses two strategies to mitigate this:

\begin{itemize}
  \item \textbf{Uniform Branching:} Where possible, conditionals are
  designed so that threads in a warp make consistent decisions.
  \item \textbf{Predication and Simplified Conditionals:} For short
  conditional regions, code is predicated to avoid divergent branching
  altogether.
\end{itemize}

These strategies mirror \nrpy's CPU-oriented \simd optimizations and help
maintain high throughput on GPUs.

\subsection{Data Transfer Optimization}

Data transfers between the host and device occur over relatively slow
buses, making them a potential bottleneck if not handled efficiently.
\nrpy addresses this in the following ways:

\begin{itemize}
  \item \textbf{Hybrid CPU+GPU Work Distribution:} By performing only
  large, data-intensive parts of the simulation on the GPU, \nellg avoids
  repeated data transfers for small workloads. \algref{alg:hybriddriver}
  outlines our approach, which ensures that tasks remain on the CPU if
  the overhead of transferring data to the GPU and scheduling a kernel
  would exceed any potential speedup.
  \item \textbf{Pinned Memory and Asynchrony:} \nrpy allocates
  \textit{pinned} (page-locked) memory on the host using
  \texttt{cudaMallocHost}, facilitating faster, asynchronous
  \textit{host--device} transfers. Critical housekeeping tasks (e.g.,
  computing the grid $L^2$ norm of the Hamiltonian constraint violations)
  are overlapped with data transfers so that the GPU remains busy while
  data is being moved.
  \item \textbf{\cuda Streams:} Multiple \cuda streams can be used to
  schedule concurrent kernel executions and asynchronous copies. In
  multi-patch or multi-grid contexts, streams help overlap computations
  for different patches, although the best performance gain is achieved
  when each kernel is sufficiently large to hide scheduling overheads. By
  default, we set the number of streams to $\texttt{nstreams} = 3$, one
  per coordinate direction, but we find only a marginal speed-up using
  more than one stream and an insignificant speedup for
  $\texttt{nstreams} > 3$. In other scenarios or more compute heavy
  kernels, streams may prove to be more beneficial.
\end{itemize}

\subsection{\cuda Intrinsics}

A key advantage of \nrpy is its ability to generate explicitly vectorized
code, combining common subexpression elimination (CSE) with hardware
intrinsics to aggressively fuse arithmetic operations. This becomes
increasingly important for NR applications as the mathematical
expressions get so long that it can be challenging for compilers to
optimize the code effectively. By default, \nrpy detects long arithmetic
expressions in finite-difference stencils and replaces repeated
operations with corresponding \cuda intrinsics where appropriate,
reducing the total number of floating-point operations in the final
compiled kernel.  Intrinsics in the \cuda setting (e.g.,
\texttt{\_\_dmul\_rn}, \texttt{\_\_dadd\_rn}, or \texttt{\_\_fma\_rn})
ultimately results in fewer instructions being executed and more efficient
cache usage, thus improving performance and reducing rounding errors\footnote
{We utilize intrinsics based on the ``round to nearest even'' rounding mode.}
.

\begin{algorithm}[h]
  \caption{\nellg Driver: \textbf{Host} (\textbf{Device}) denotes a task
  performed on the CPU (GPU). The most computationally expensive
  operations are boldfaced: \textbf{H}, \textbf{RHS}, \textbf{BC}, and
  \textbf{RK}.}
  \label{alg:hybriddriver}
  \raggedright
  \begin{algorithmic}[1]
    \State \textbf{Host}: Initialize global array of \cuda streams.
    \State \textbf{Device}: Set up uniformly sampled coordinate 1D arrays
    $x^i$, transfer to \host.
    \State \textbf{Device}: Precompute reference metric components and
    derivatives.
    \State \textbf{Host}: Initialize ``inner'' and ``outer'' boundary
    conditions~\cite{Ruchlin:2017com,Assumpcao:2021fhq} containers,
    transfer to \device.
    \State \textbf{Device}: Allocate storage for Runge-Kutta stages and
    constant source terms grid functions.
    \State \textbf{Host}: Allocate storage for diagnostics stored on the
    entire grid.
    \State \textbf{Device}: Set initial conditions and compute constant
    source terms.

    \While{$t \leq t_{\rm final}$}
      \State \textbf{Device}: Compute residual (\textbf{H}; left-hand side of
      Eq.~\ref{eqn:hamiltonian_constraint_v3})
      \State \textbf{Host}: Request asynchronous data transfer from
      \device for diagnostics.
      \State \textbf{Device}: Compute residual $L^2$ norm.
      \While{Runge-Kutta step incomplete}
        \State \textbf{Device}: Evaluate right-hand sides (\textbf{RHS})
        of Eqs.~\ref{eq:two_punctures_rfm_system}.
        \State \textbf{Device}: Apply boundary
        conditions~\cite{Assumpcao:2021fhq} (\textbf{BC}) to evolved
        variables $u$ and $v$.
        \State \textbf{Device}: Perform Runge-Kutta substep (\textbf{RK})
        update.
      \EndWhile

      \State \textbf{Host}: Compute timestep.
      \State \textbf{Host}: Check alternate stop condition based on the
      $L^2$ norm of residual.
    \EndWhile
    \State Synchronize \device and \host.
    \State Free \device and \host allocated storage.
    \State Program terminates.
  \end{algorithmic}
\end{algorithm}

\subsection{Scalability}

The flattened array layout and well-structured kernel design allow
\nrpy-based codes to scale effectively across GPUs ranging from
consumer-grade (NVIDIA RTX series) to HPC-grade (A100, L40, etc.). Since
each GPU has more streaming multiprocessors (SMs) than a CPU has cores,
the large and compute-heavy kernels generated by \nrpy often achieve
near-peak bandwidth usage, as shown in Sec.~\ref{sec:results}. This
approach ensures that as problem sizes grow or as more advanced hardware
becomes available, \nellg remains a viable solution for numerically
challenging NR applications.

\subsection{Complete \nellg Driver Algorithm}

The complete \nellg driver workflow is detailed in
\algref{alg:hybriddriver}. The solver begins by allocating and
initializing data structures on both the host (CPU) and device (GPU). It
then enters the primary relaxation loop, which executes on the GPU, while
the host manages auxiliary tasks. Convergence checks and diagnostic
computations are performed asynchronously to optimize performance. Once a
stopping criterion is satisfied, \nellg synchronizes operations and
deallocates resources on both the host and device before terminating the
execution.

With these core optimizations in place, we now turn to the performance
benchmarks and accuracy studies of \nellg, which confirm both its
consistency with the CPU-based \nell code and its ability to deliver high
performance across various GPU platforms.

\section{Results}
\label{sec:results}
\begin{table}[t]
  \centering
  \caption{\label{table:system_hardware} Specifications of the hardware
  tested, including a consumer-grade PC and a standard node in the
  \texttt{Falcon} cluster. Each \texttt{Falcon} node contains two CPUs.
  Finally, we note the capacity and bandwidth of dynamic random
  access memory (DRAM) and thermal design power (TDP).}
  \vspace{0.5em}
  {
    \footnotesize
    \setlength{\aboverulesep}{0pt}
    \setlength{\belowrulesep}{0pt}
    \begin{tabular}{l!{\vrule}cd{3.0}d{3.3}c}
      \toprule
      \multicolumn{1}{c!{\vrule}}{\multirow{2.25}{*}{\textbf{System}}} & \multirow{2.25}{*}{\textbf{Model}}          & \multicolumn{2}{c}{\textbf{DRAM}} & \multirow{2.25}{*}{\textbf{TDP (W)}} \\
      \cmidrule(lr){3-4}
      & & \multicolumn{1}{c}{\textbf{Capacity (GB)}} & \multicolumn{1}{c}{\textbf{Bandwidth (GB/s)}} & \\[0.4ex]
      \midrule\\[-2.35ex]
      \texttt{Desktop-CPU} & Ryzen 9 5950x  & 64 & 42.7 & 105 \\
      \texttt{Desktop-GPU} & RTX3080 & 12 & 912.0 & 320 \\
      \texttt{Falcon-CPU}  & Xeon E5-2695v4 & 128 & 76.8 & 120 \\
      \texttt{Falcon-GPU}  & L40            & 40 & 864.0 & 300 \\
      \bottomrule
    \end{tabular}
  }
\end{table}

Using the hardware described in Sec.~\ref{ssec:hardware}, we present four
studies. First, Sec.~\ref{ssec:consistency} compares \nellg and \nell to
verify that our \cuda implementation achieves numerical accuracy
consistent with the trusted \openmp version, agreeing at roundoff levels.
Second, Sec.~\ref{ssec:roofline} evaluates the weak algorithmic scaling
of the core computational kernels introduced in
\algref{alg:hybriddriver}: \textbf{RHS} (Right-Hand Side), \textbf{BC}
(Boundary Conditions), \textbf{RK} (Runge-Kutta substeps), and \textbf{H}
(Hamiltonian Constraint). Third, Sec.~\ref{ssec:intrinsics} investigates
the impact of intrinsics on performance and accuracy. Finally,
Sec.~\ref{ssec:scalability} examines the scalability of \nrpy-generated
GPU kernels for HPC systems and \bhah's multipatch grids, demonstrating
their suitability for larger-scale simulations.

\subsection{Hardware Overview}
\label{ssec:hardware}
Except for Fig.~\ref{fig:falcon_results},%
\footnote{See Sec.~\ref{ssec:intrinsics} for details.}
all results were obtained on a single consumer desktop (see
Tab.~\ref{table:system_hardware} for specifications). This desktop
includes an AMD Ryzen 9 5950x (CPU) and a NVIDIA RTX3080 (GPU) with
compute capability $8.6$. Each implementation employs $10$th-order
finite-difference stencils. Comparisons with \nell use its highly
optimized, \openmp-parallelized version, which benefits from \nrpy's
\simd optimizations.  Here we restrict \openmp to one thread per
\textit{physical} core, as no noticeable performance benefit was measured
when using hyperthreads since the available cache per thread is reduced.

\subsection{Consistency Study: Roundoff-Level Agreement between \nell and \nellg}
\label{ssec:consistency}
\begin{figure}[t]
  \raggedleft
  \includegraphics[width=0.85\columnwidth]{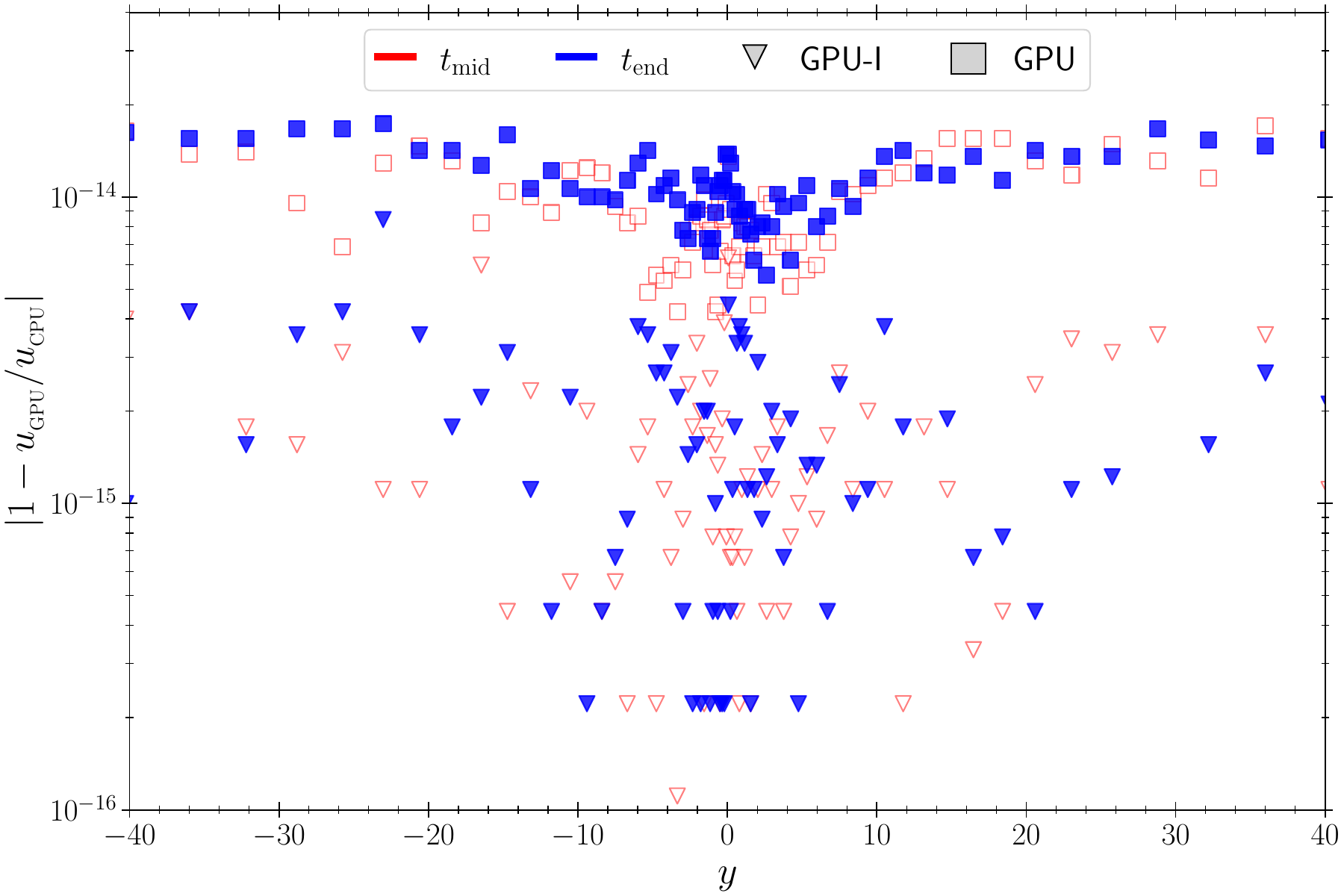}
  \caption{
    Solution comparison between \nell and \nellg halfway through
    relaxation (red) and at the end of relaxation (blue). Triangle (square)
    markers denote generating \nellg with (without) \cuda intrinsics.}
  \label{fig:solution_cmp}
\end{figure}

To establish consistency, we verify that solutions from \nellg and \nell
agree at roundoff levels. Figure~\ref{fig:solution_cmp} displays the
relative difference in the solution $u$ along grid points nearest to the
$y$-axis. Red squares represent the comparison halfway through relaxation
($t_{\rm mid}$), while blue squares depict it at the end of relaxation
($t_{\rm end}$). Both solutions are computed on a \GridRes{128}{128}{16}
grid using \nrpy's \SinhSymTP coordinate system (see \pone\ for details).
The solutions show excellent agreement, with a norm of approximately
$\num{9e-13}$, indicating minimal, roundoff-level discrepancies. As we'll
find in Sec.~\ref{ssec:intrinsics}, enabling intrinsics can further
reduce these discrepancies.

\subsection{Efficiency Study: Roofline Analysis}
\label{ssec:roofline}
Having established consistency with the trusted \nell code, we evaluate
the efficiency of \nell and \nellg on the CPU and GPU, respectively. For
profiling, we use \textit{Likwid~5.3} on the CPU and \textit{NVIDIA
Nsight Compute~2022.3.0.0} on the GPU, focusing on the four most
computationally intensive kernels: \textbf{RHS}, \textbf{H}, \textbf{RK},
and \textbf{BC}.

\subsubsection{Roofline Analysis Methodology}

For our roofline analysis, we plot the number (billions) of
floating-point operations per second (GFLOP/s) versus the arithmetic
intensity (AI; FLOP/Byte). The lower ``roof'' represents memory
bandwidth, and the upper ``roof'' corresponds to the theoretical peak
FLOP/s. Although our analysis focuses on main memory bandwidth
(DDR/GDDR), its principles extend to various cache levels. It is
important to emphasize that AI is strongly tied to the
memory demand of a kernel as well as the complexity of the calculation.
Stated differently, a sufficiently complex calculation can outweigh the
memory access latency if there is enough work to be performed.

\begin{figure}[t!]
  \raggedleft
  \includegraphics[width=0.85\columnwidth]{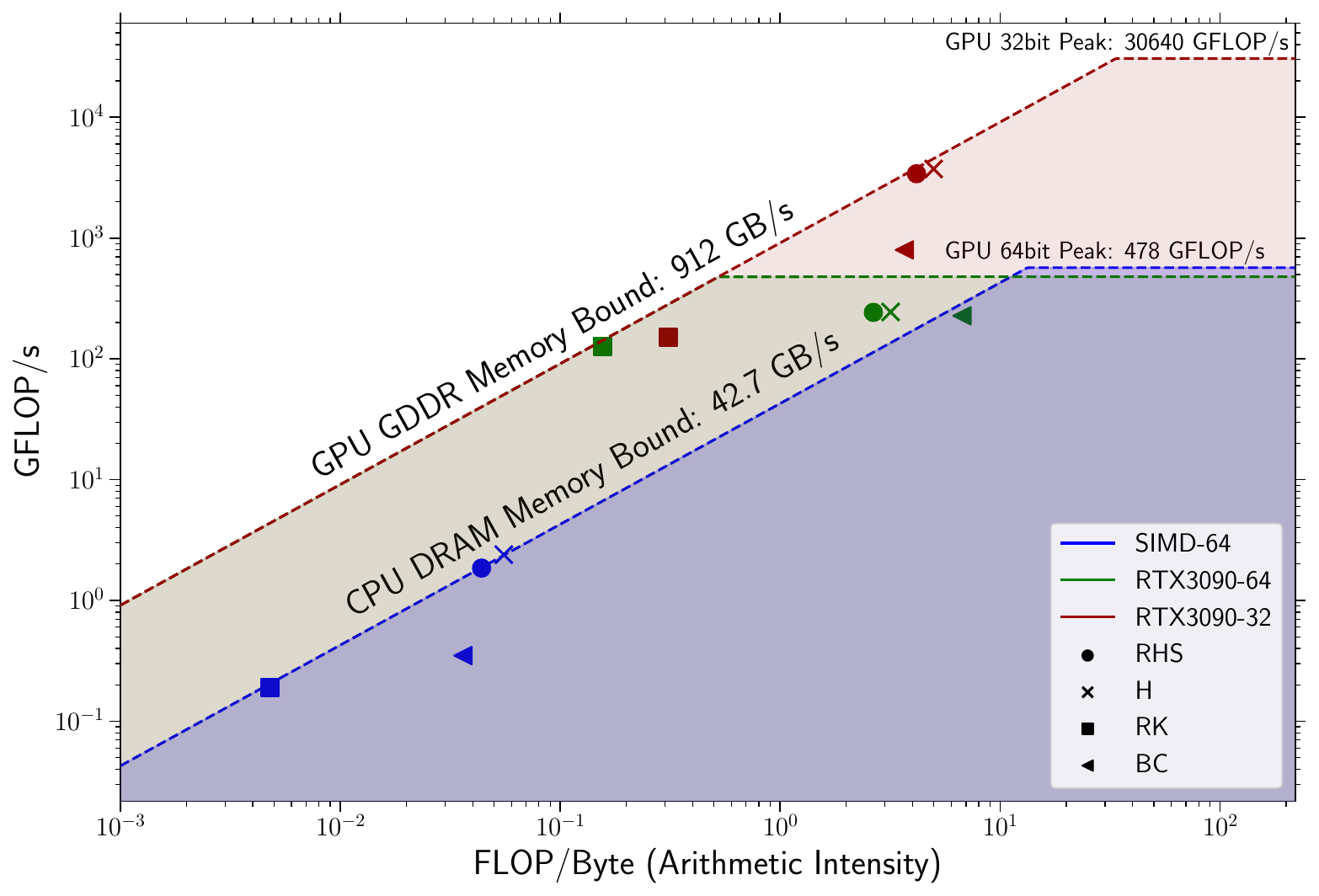}
  \caption{Roofline comparison of the vectorized (\simd) CPU version of
    \nell and the accelerated \nellg (GPU) codes. Here we plot the data
    for the \textbf{RHS} (Right-Hand Side), \textbf{H} (Hamiltonian Constraint), \textbf{RK} (Runge-Kutta substeps), and \textbf{BC} (Boundary Conditions) kernels. CPU metrics
    were obtained using \textit{Likwid 5.3}, while GPU metrics were
    obtained using \textit{NVIDIA Nsight Compute 2022.3.0.0}.}
  \label{fig:roofline}
\end{figure}

\subsubsection{Performance Metrics and Observations}

Figure~\ref{fig:roofline} compares CPU (dashed blue for double precision)
and GPU (dashed red for single precision, dashed green for double
precision) performance on a \GridRes{512}{512}{64} grid in \SinhSymTP
coordinates. Here we denote the kernel performance for the \textbf{RHS}
(circles), \textbf{H} (crosses), \textbf{RK} (squares), and \textbf{BC}
(triangles). The first observation to note is that the CPU's
double-precision peak performance slightly surpasses that of the GPU.
Additionally, the GPU's single-precision peak is approximately 64x higher
than its double-precision peak, reflecting the optimization of
consumer-grade GPUs for single-precision performance. Finally, the elbows
of the roof (i.e., the transition to the upper roof) denote the threshold
from an algorithm that is memory bound (above the roofs) to one that is
compute bound (below the roofs).

Focusing first on the CPU results (blue markers), we find that all
kernels are heavily memory bound, resulting in low AI ($10^{-3} \lesssim
{\rm AI} < 10^{-1}$), with RK being the lowest. Conversely, GPU kernels
generally achieve higher AI ($10^{-1} \lesssim {\rm AI} < 10^{1}$), with
GFLOP/s near the GPU's peak double precision performance.  The GPU
kernels are primarily compute-bound except for \textbf{RK}, which
remains memory-bound due to its minimal arithmetic workload. The GPU's
higher AI is largely attributed to its 21x greater memory bandwidth and
the GPU's ability to significantly hide latency by having considerably
more active threads executing instructions each clock cycle. Furthermore,
by moving as much information to compile time regarding the memory
layout, access patterns, and efficient use of \cuda constant cache, the
\cuda compiler is able to effectively optimize memory accesses.

To gain further insight into the discrepancies between single and double
precision calculations on the GPU, it is important to first identify the
inherit limitations of devices with compute capability 8.6. Specifically,
these devices can perform at most 2 double-precision calculations per
clock cycle, while up to 128 single-precision calculations can be
performed per clock cycle (see section 5.4.1 of~\cite{nvidia_cpg}).
Therefore, for an optimized kernel with sufficiently high complexity
(i.e., ${\rm AI} > 10^1$), the achieved FLOP/s in single-precision should
be 64x more than for double precisions.

To this end, we have implemented strong floating-point typing into \nrpy
to enable the generation of optimized single-precision executables. Here
we have leveraged this capability to generate the single-precision
version of \nellg and have included the associated roofline results in
Fig.~\ref{fig:roofline}.%
\footnote{We have verified with \textit{NVIDIA Nsight Compute} that there
are no double-precision calculations counted for the RK, RHS, and RK
kernels.}
We find that the AI is roughly constant with the achieved FLOP/s being
${\sim}10{\rm x}$ higher than for double-precision, at which point
the H and RHS kernels become memory bound. Therefore, the
single-precision kernels are not able to achieve the theoretical maximum
speedup of 64x.

\begin{figure}[t!]
  \raggedleft
  \includegraphics[width=0.85\columnwidth]{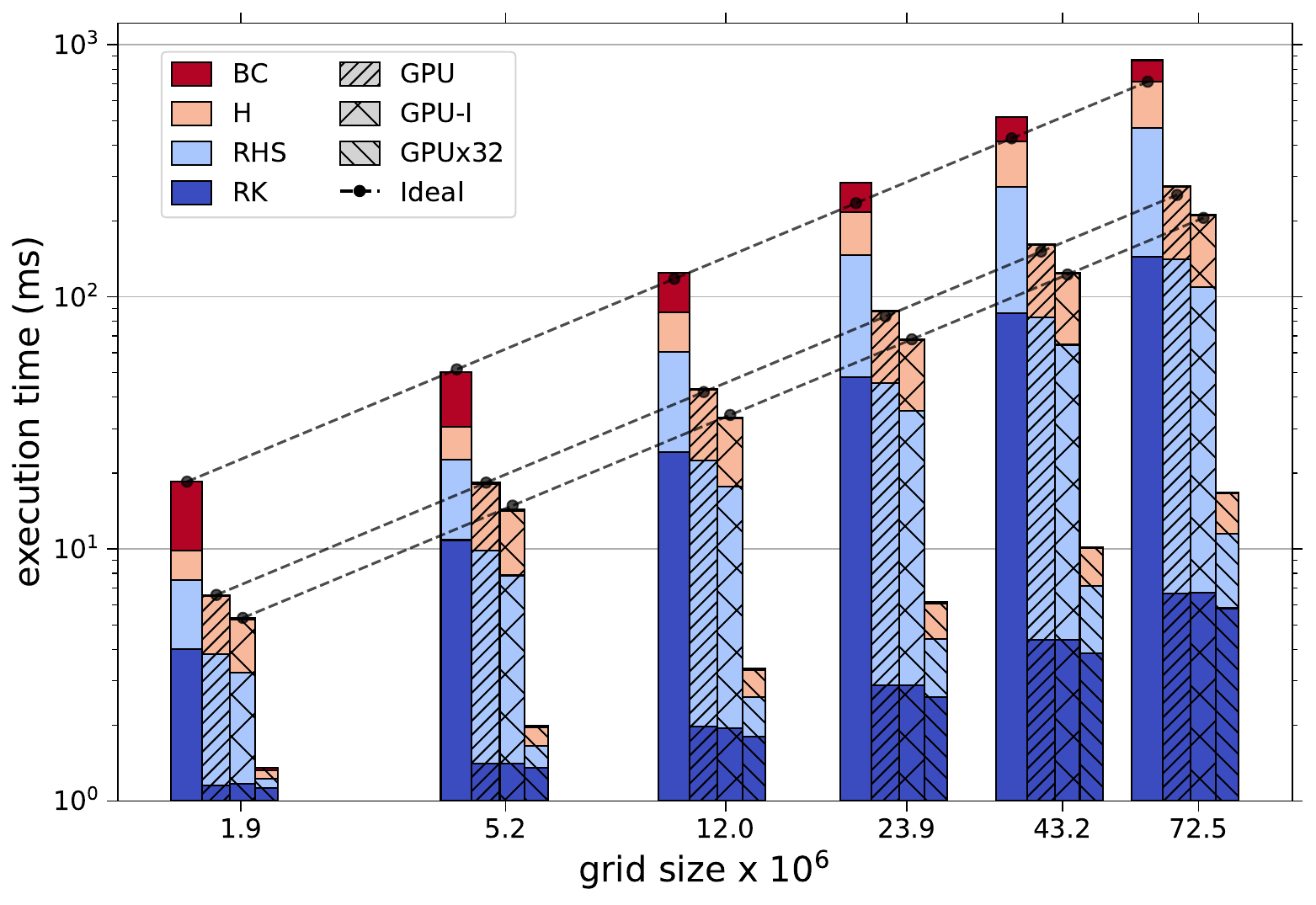}
  \caption{For each bar, we show the execution time for a single call to
    each kernel, not the entire program runtime, for increasing grid
    sizes. We compare the \nell CPU code (no hatch marks) against
    \nellg without \cuda intrinsics (GPU), \nellg with \cuda intrinsics
    (GPU-I), and \nellg using single precision (GPU$\times$32). Dashed lines
    denote approximate ideal weak scaling.}
  \label{fig:weak_scaling}
\end{figure}

In Fig.~\ref{fig:weak_scaling}, we illustrate the above speed-ups using
weak algorithmic scaling of the most computationally intensive kernels,
\textbf{BC} (red), \textbf{H} (light red), \textbf{RHS} (light blue), and
\textbf{RK} (blue), and compare the accumulated execution time for
increasing grid sizes. Both \nell and \nellg show effective parallelization
given they match well against ideal scaling estimates (dashed lines).

\subsubsection{Comparative Analysis with Other GPU-Enabled NR Codes}

Direct comparisons with other GPU-enabled NR codes are nontrivial;
however, to gauge the effectiveness of our implementation, we contrast our
roofline analysis against results from previous literature for the \cuda
port of \texttt{Dendro-GR}~\cite{Fernando:2022zbc} and the early \kokkos
port of \texttt{Athena}, \texttt{K-Athena}~\cite{grete2020k}. Specific
roofline observations include:

\begin{itemize}
  \item \texttt{Dendro-GR}: In Fig.~14 of Ref.~\cite{Fernando:2022zbc},
  the AI for \textit{octant-to-patch} operations ($m_i$) aligns well with
  our results. However, their RHS kernel exhibits ${\rm AI} < 10^0$ even
  on a higher-end NVIDIA A100 (compute capability $8.0$), largely
  associated with cache misses and register spillage inherent to solving
  the full system of Einstein's equations in 3+1 form, which are far more
  complex than Eqs.~(\ref{eq:two_punctures_rfm_system}).

  \item \texttt{K-Athena}: In Fig.~2 of Ref.~\cite{grete2020k}, the
  reported AI for the 3D linear wave problem is approximately 1.5 on an
  NVIDIA V100 (compute capability 7.0), which is only ${\sim}2{\rm x}$
  higher than the reported CPU AI.
\end{itemize}

We note that both references use data center grade GPUs, where each SM is
capable of computing $32$ double-precision calculations per clock cycle
as compared to the GPUs used in this work which are restricted to $2$
double-precision calculations per clock cycle.

Overall, we conclude that \nellg demonstrates high efficiency on
consumer-grade GPUs, with the potential for further speedups when used
with modern data center-grade GPUs with considerably higher
double-precision throughput.

\subsection{Impact of Intrinsics}
\label{ssec:intrinsics}
We next assess the effect of intrinsics, i.e., specialized \cuda
instructions (e.g., fused multiply-add), that can reduce total
instructions, thus improving efficiency. To quantify this, we compare the
executed instruction counts and categories using \textit{NVIDIA Nsight
Compute}, both without (``No intrinsics'') and with (``Intrinsics'')
\cuda intrinsics when generating the \textbf{RHS} kernel.

\begin{figure}[t!]
  \raggedleft
  \includegraphics[width=0.85\columnwidth, keepaspectratio]{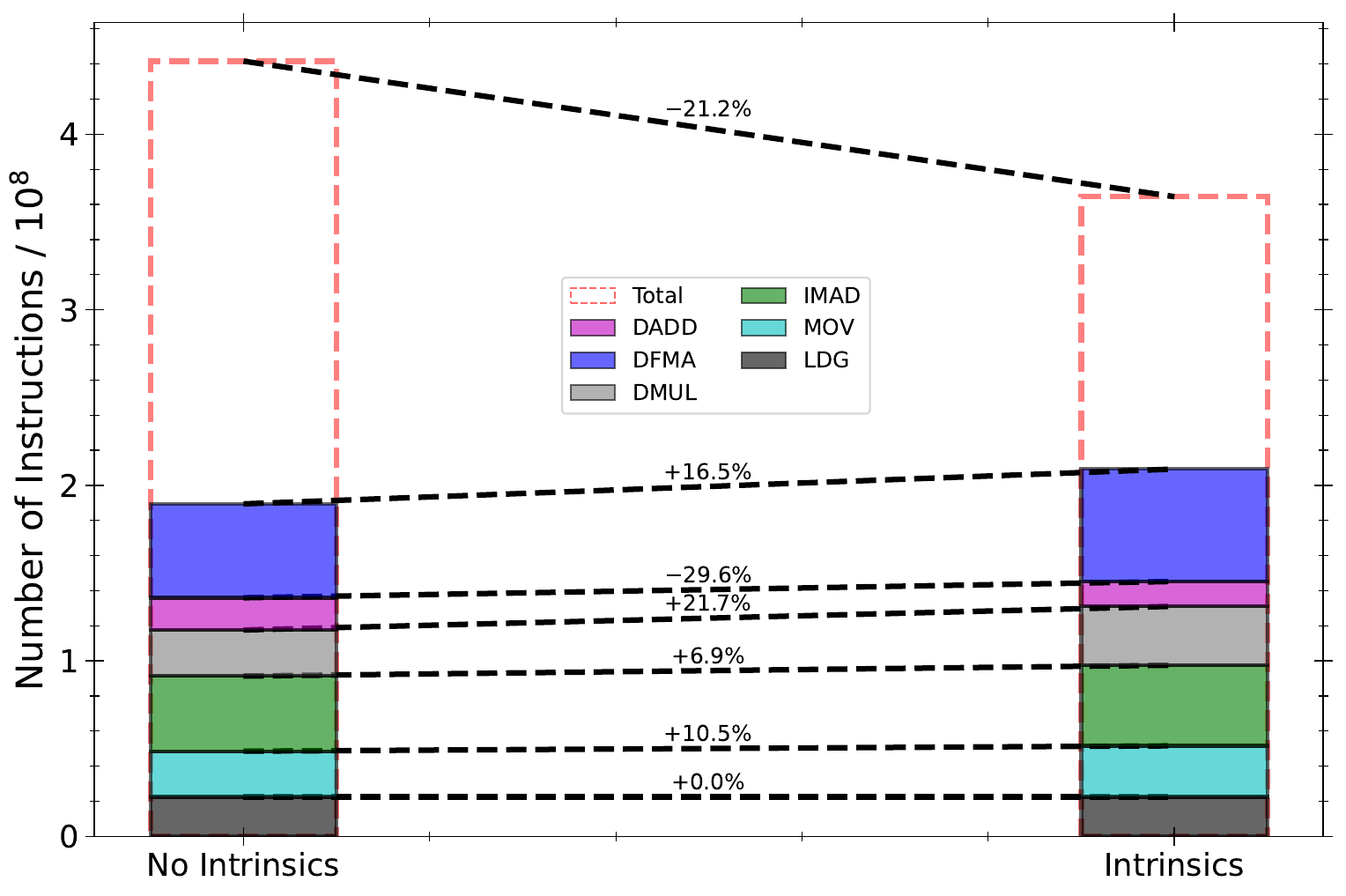}
  \caption{Instruction distribution for the \nellg \textbf{RHS} kernel,
    comparing the code port without intrinsics (No intrinsics) to one
    using \cuda intrinsics (Intrinsics). The ``Total Instructions'' bar
    shows the overall reduction in instruction count.}
  \label{fig:instr_cmp}
\end{figure}

Figure~\ref{fig:instr_cmp} demonstrates that enabling intrinsics reduces
the total executed instructions by approximately \qty{21}{\percent}. This
is largely attributed to the ${\sim}\qty{16.5}{\percent}$ increase in
DFMA (Double-Precision Fused Multiply-Add), a key component to reducing
DADD (Double-Precision Add) operations by ${\sim}\qty{29.6}{\percent}$.
We further find a ${\sim}\qty{10.5}{\percent}$ increase in MOV
operations, which implies more efficient cache use during calculations.
Enabling intrinsics in the \textbf{H} and \textbf{RHS} kernels further
improves total runtime by 1.3x and 1.2x, respectively, compared to the
No-Intrinsics \nellg port and by ${\sim}4{\rm x}$ relative to \nell
(Fig.~\ref{fig:weak_scaling}, GPU-I). However, when energy usage is
estimated based on TDP-per-unit-speedup, the energy efficiency gains are
more modest, yielding only a 1.3x improvement over \nell\footnote
{Note: the CPU power usage during execution of the GPU application is not
considered as a robust method to determine the CPU and GPU power usage at
runtime was not found.  Furthermore, using TDP assumes each device is
functioning at their peak power at runtime, which is not necessarily the
case.}.

An unexpected advantage of adopting intrinsics is slightly improved
numerical agreement with \nell as shown in Fig.~\ref{fig:solution_cmp}
(triangles), where the discrepancy between solutions decreases by
$10^1$--$10^2$. Thus, enabling intrinsics enhances both performance and
accuracy.

\subsection{Scalability}
\label{ssec:scalability}
The flattened array layout and structured kernel design enable \nellg to
scale effectively across single GPUs, from consumer-grade (e.g., RTX
series) to HPC-grade (e.g., A100, L40). With significantly more streaming
multiprocessors (SMs) than CPU cores, GPUs allow \nellg to saturate
bandwidth for its larger kernels, as demonstrated in
Sec.~\ref{ssec:roofline}. We conclude that the \cuda kernels emitted by
\nrpy provide a highly performant foundation for \nellg and pave the way
for full NR evolution codes that better exploit GPU capabilities.

\begin{figure*}[t]
  \begin{center}
    \includegraphics[width=0.5\textwidth, keepaspectratio]{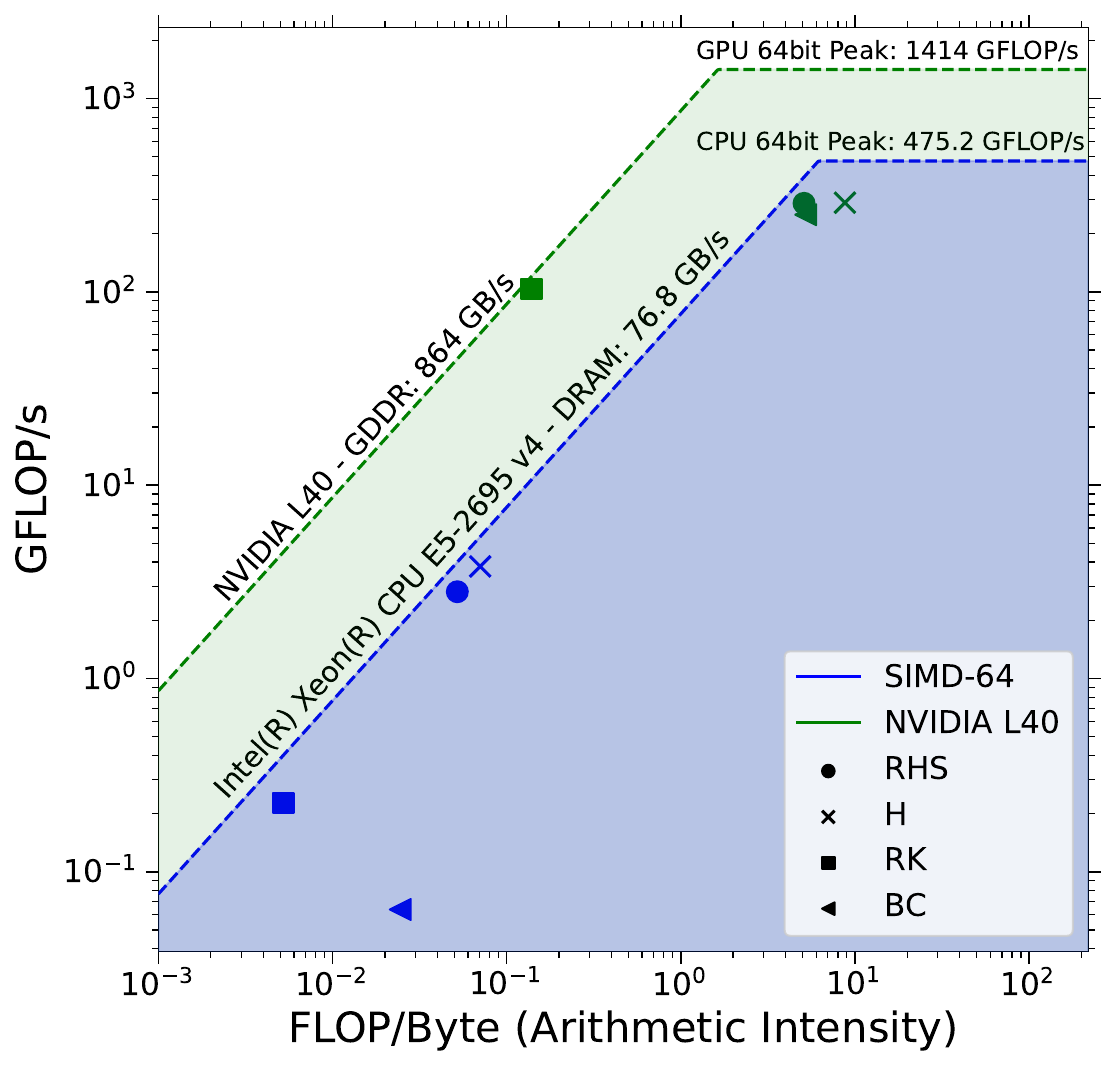}
    \includegraphics[width=0.49\textwidth, keepaspectratio]{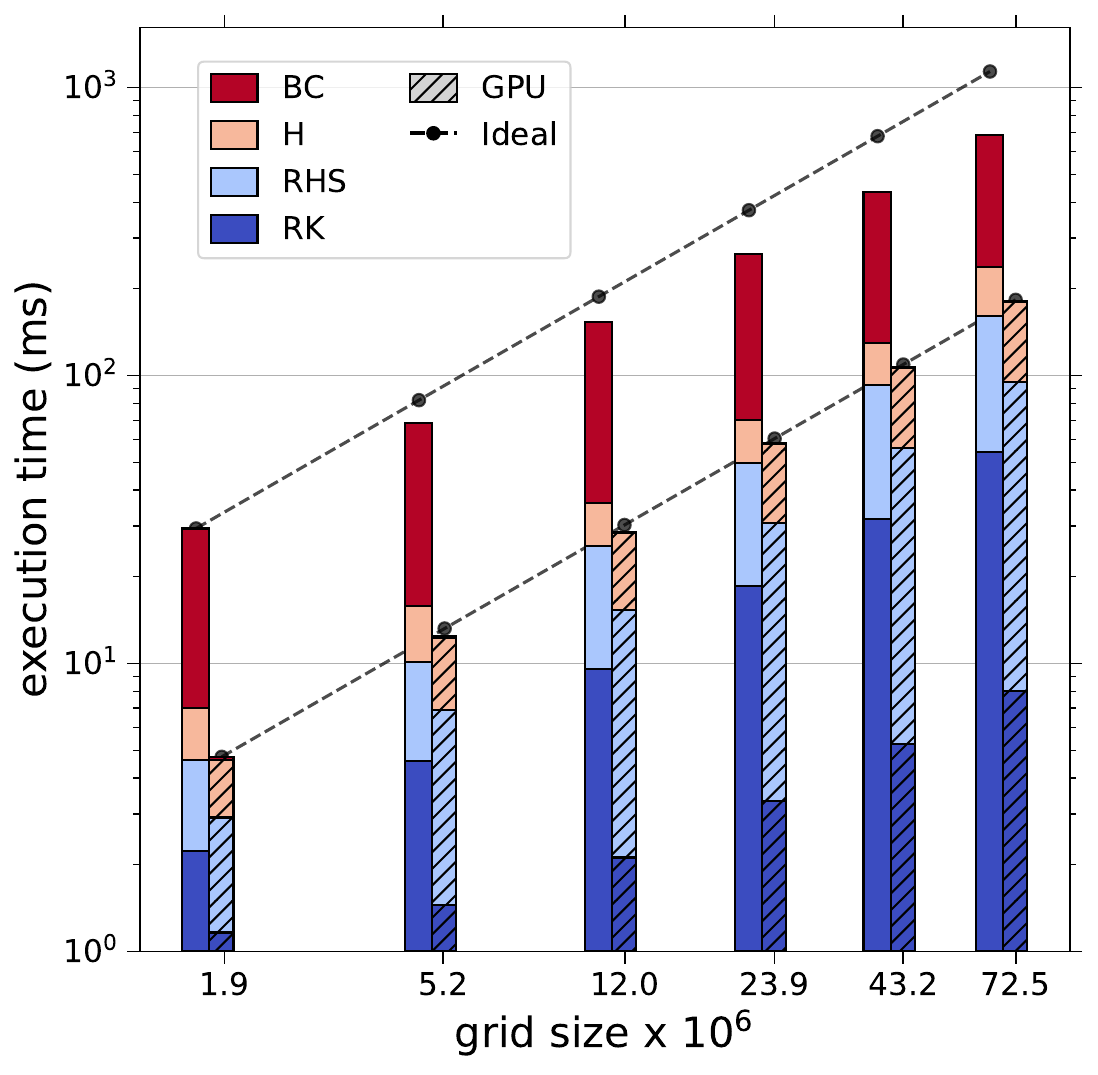}
  \end{center}
  \caption{Same as Fig.~\ref{fig:roofline} (\textit{Left}) and
    Fig.~\ref{fig:weak_scaling} (\textit{Right}), but using the HPC
    \texttt{Falcon} cluster. The L40 has compute capability 8.9, which is
    still limited to two double-precision operations per clock cycle.}
  \label{fig:falcon_results}
\end{figure*}

\subsubsection{Performance on HPC Hardware}
To gauge the performance gap between consumer and HPC hardware, we repeat
the analysis shown in Fig.~\ref{fig:roofline} and
Fig.~\ref{fig:weak_scaling} on a single node of the retired Idaho
National Laboratory cluster, \texttt{Falcon}, the results of which are
illustrated in Fig.~\ref{fig:falcon_results}. Each \texttt{Falcon} node
features two Xeon E5-2695v4 CPUs and an L40 GPU (see
Tab.~\ref{table:system_hardware} for specifications). While the L40
offers higher theoretical performance, practical gains over the RTX3080
are limited to ${\sim}$\qty{15}{\percent}. We attribute this,
potentially surprising, minimal increase in performance to their compute
capability (8.9 for the L40 and 8.6 for the RTX3080), which implies they
are both limited to two double-precision calculations per clock
cycle~\cite{nvidia_cpg}, thus further emphasizing the hardware imposed
limitations to the measured performance.

\subsubsection{Multiple, Independent Patches Performance}
\label{ssec:streams}

Finally, in preparation for \bhah multipatch grids, we performed
identical relaxations on multiple independent grids in parallel using
\nell and \nellg with $1 \le N \le 7$ identical grids of size
\GridRes{128}{128}{16}, the smallest grid size (\num{1.3e6}) used in
Fig.~\ref{fig:weak_scaling}. Furthermore, there is no inter-grid
interpolation or data sharing, therefore this purely looks at
computational efficiency without the overhead of communication. We have
also disabled diagnostic outputs during runtime except when saving the
solution at the end of the calculation to further highlight the
efficiency of the multi-grid calculations.

\begin{figure}[t!]
  \raggedleft
  \includegraphics[width=0.485\columnwidth]{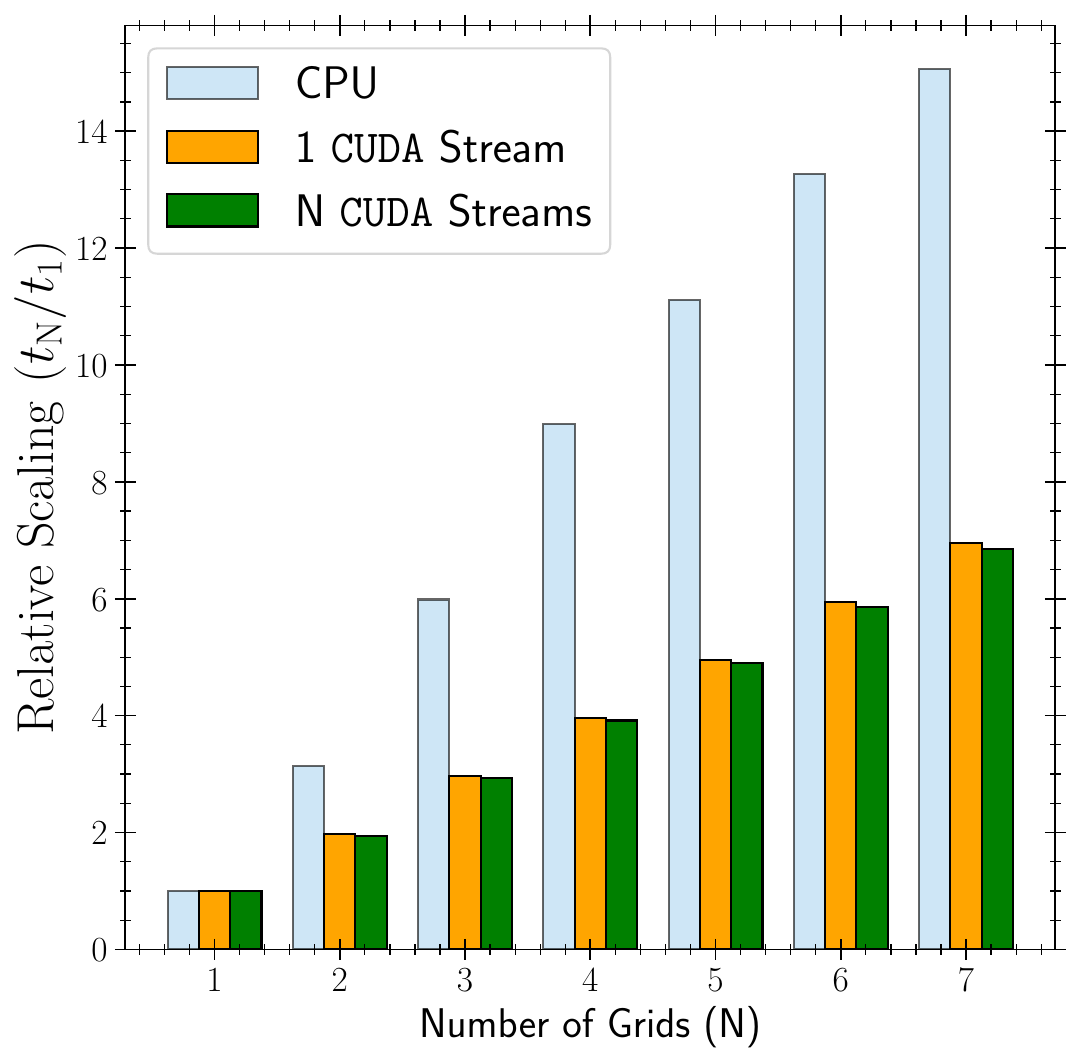}
  \includegraphics[width=0.49\columnwidth]{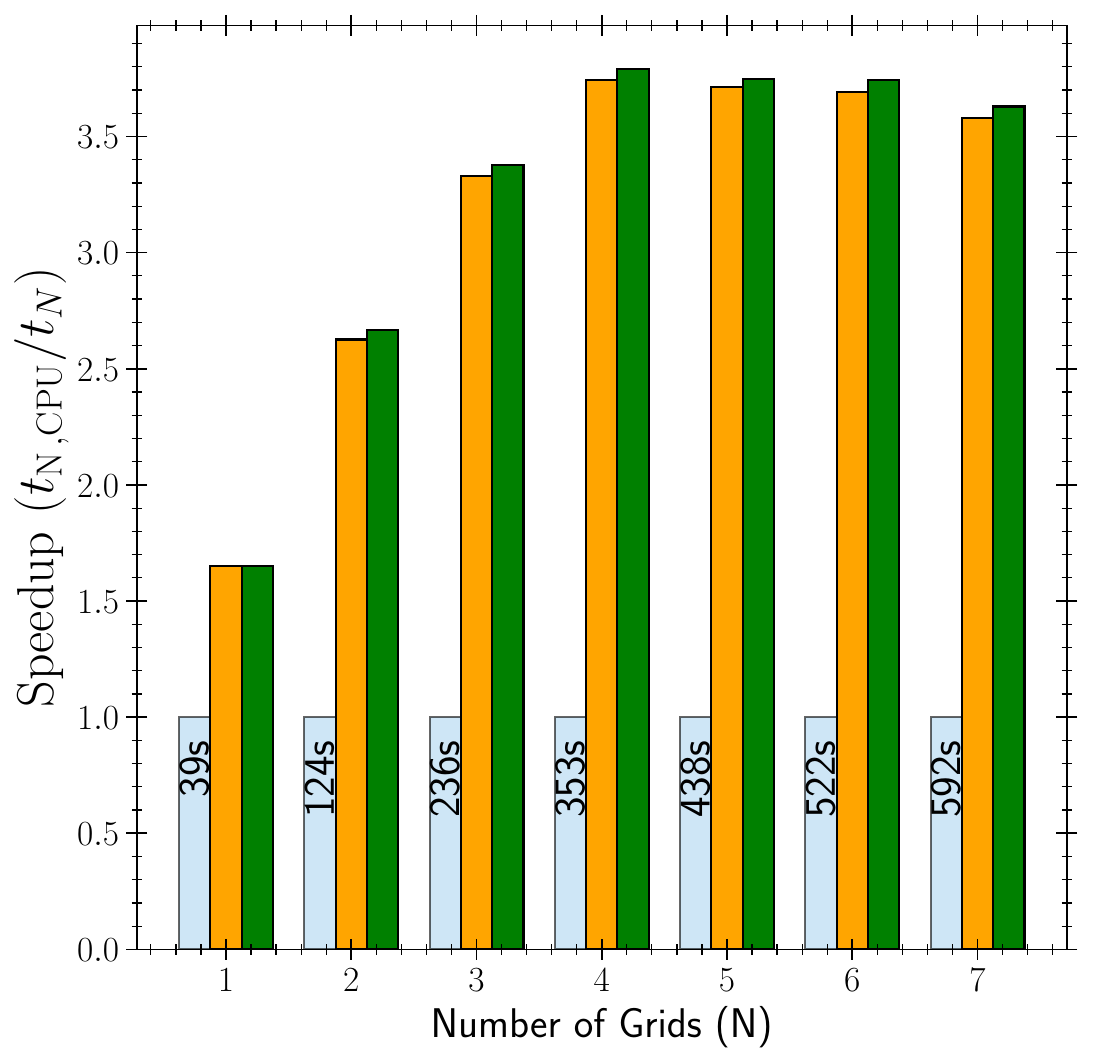}
  \caption{\textit{Left}: Total execution time ($t_{N}$) for $N$ grids,
    normalized by the single-grid runtime ($t_{1}$). \nell results (blue)
    increase faster than the nearly linear \nellg results using $1$ \cuda
    stream (orange) or N \cuda streams (green). \textit{Right}: Overall
    GPU speedup compared to \nell.}
  \label{fig:multigrid_cmp}
\end{figure}

In Fig.~\ref{fig:multigrid_cmp} (left) we illustrate the ratio of the
total runtime for a given number of grids to the total runtime for a
single grid, where results for \nell are in blue and results for \nellg
with intrinsics using $1$ (N) \cuda streams are shown in orange (green).
Here we find that \nellg scales nearly as $\mathcal{O}(N)$, whereas \nell
exhibits scaling closer to $\mathcal{O}(N^{1.5})$.

To gauge the relative speedup of \nellg vs \nell, we illustrate in
Fig.~\ref{fig:multigrid_cmp} (right) the speedup as a function of the
number of grids. We define the speedup as the total CPU runtime to
solution for a given number of grids ($t_{\rm N \,, CPU}$) normalized by
$t_{\rm N}$ for CPU and GPU execution. For the CPU, the speedup bar is
always $1$, included for clarity and to annotate $t_{N \,, \rm CPU}$ for
each CPU bar. For \nellg, the measured speedup ranges from a minimum of
1.65x for a single coarse grid to a maximum of 3.79x with four grids,
averaging 3.23x. This behavior suggests the scheduler is likely saturated
at four grids, with the decline for ${N} > 4$ likely due to launch
latency overhead.

These benchmarks were repeated using $\nstreams \in \left\lbrace 1 \,, 3
\,, N \right\rbrace$ to evaluate the benefits of additional \cuda
streams. Using $\nstreams = N$ is approximately 1.2x faster than
$\nstreams = 1$ and at most 0.5x faster than using $\nstreams = 3$,
regardless of $N$. The marginally maximum benefit occurs when $\stream =
N = 4$, reinforcing that saturation of the launch scheduler at $N \sim 4$
significantly contributes to overall latency. Additionally, the grid
coarseness results in a less expensive kernel, which may limit the
ability to fully benefit from multiple streams.

These findings demonstrate the ability of GPUs to manage multiple
independent patches while minimizing latency through efficient \cuda
scheduling. Such scalability is crucial for leveraging GPUs for
large-scale NR simulations. For these tests, synchronizations between the
\host and \device are minimal, limited to data transfers for disk
storage. Therefore, these results represent an important upper
performance bound, as the patches are independent and require no
inter-patch data sharing.

\section{Conclusions \& Future Work}
\label{sec:conclusions}

In this work, we extended the Python-based \nrpy code generation
framework to generate optimized \cuda-enabled programs, marking a major
step in adapting NR codes to use GPU architectures. As a first example of
this improved capability, we developed \nellg, the first GPU-accelerated
elliptic solver aimed at solving the BBH initial value problem. Using
\nrpy's flexible code generation for various coordinate systems, \nellg
retains the adaptability of its CPU-based predecessor, \nell, supporting
Cartesian-like, spherical-like, cylindrical-like, and bispherical-like
geometries.

Our tests show that \nellg produces results that match \nell at roundoff
levels, ensuring accurate solutions for NR simulations. By leveraging the
GPU's \simt model and high-bandwidth memory, \nellg shifts key
calculations from being memory-bound on CPUs to being compute-bound on
GPUs. This optimization leads to large performance gains, with \nellg
running about 4x faster on an NVIDIA RTX3080 GPU using double precision.
On HPC-grade hardware (NVIDIA L40), performance increases by only $\sim
15\%$ compared to the RTX3080, reflecting the shared limitation of two
double-precision operations per clock cycle for both architectures.
Switching to single precision provides roughly a 16x speedup for the more
computationally intensive kernels, rather than the theoretical 64x, as
the application becomes memory bound.  This suggests that \nellg using
double precision would be significantly more performant on, e.g. an
NVIDIA V100 or A100, which are capable of $32$ or $64$ double precision
calculations per clock cycle, respectively.

A roofline analysis supports these observations, demonstrating that
\nellg's kernels can achieve up to a $10^2$-fold improvement in
arithmetic intensity (AI) compared to CPU versions. This improvement
arises from more efficient memory access patterns, lower data transfer
overhead, and carefully tuned \cuda kernels (see
Fig.~\ref{fig:roofline}). Adding \cuda intrinsics---specialized
instructions that fuse arithmetic operations---reduces instruction counts
by approximately \qty{21}{\percent} for certain kernels (e.g., \textbf{H}
and \textbf{RHS}), resulting in an additional $1.2$--$1.3{\times}$
speedup over the non-intrinsic GPU version (and about $4\times$ relative
to \nell). Intrinsics also improve numerical agreement with \nell by one
to two orders of magnitude.

\nellg's algorithmic design minimizes communication overhead between the
\host and \device, limiting synchronizations to diagnostics during the
hyperbolic relaxation procedure. Local coordinate storage and
asynchronous data transfers ensure smooth data movement within
single-grid applications. These optimizations have provided valuable
insights for future work. Extending these methods to multi-patch
simulations and solving Einstein's equations in full will require
tackling similar challenges, along with managing the greatly increased
register pressure associated with larger kernels, which could
significantly impact performance.

To gauge the efficiency of \nellg, we have compared with other
GPU-enabled NR frameworks, such as \texttt{Dendro-GR} and
\texttt{K-Athena}, which indicates that \nellg achieves competitive
performance despite its focus on single-grid applications and simpler
systems of PDEs. We note that direct comparison is not possible,
especially since these frameworks support adaptive mesh refinement and
more complex physics.  However, \nellg's ability to handle
computationally demanding tasks with reduced memory bottlenecks
underscores the benefits of \nrpy's automatic code generation for
specialized high-performance kernels and its potential for future
multi-grid applications.

Looking ahead, our \nrpy-based \cuda extensions are designed to integrate
seamlessly into full NR evolution codes (e.g., \bhah), unlocking the
potential of both consumer- and HPC-grade GPUs for large-scale BBH
simulations. Several complex tasks to achieve these goals include
efficiently parallelizing interpolation between grids and finding an
optimal GPU kernel adaptation for the BSSN system which has proven to be
challenging~\cite{Fernando:2022zbc}. Additionally, this work provides a
template that can be used to extend \nrpy to additional architectures
(e.g. HIP), thus removing the restriction to \cuda enabled devices.
Collectively, these developments could enable the crowd-sourced
generation of extensive GW catalogs and facilitate the exploration of
multi-messenger phenomena within NR.
Finally, we also plan to incorporate GPU acceleration into our
\texttt{Charm++}-capable version of \bhah, enabling efficient use of
multi-GPU setups and HPC resources. This extension would open up regions
of BBH parameter space that are beyond the reach of consumer-grade
hardware. It will also address challenges such as efficient load
balancing and support for heterogeneous architectures.

In summary, the development and validation of \nellg underscore the
potential of GPU acceleration for NR applications and the power of code
generation using \nrpy. With significant performance gains and robust
accuracy, \nellg underscores how modern computing architectures and
automatic code generation can meet the increasing demands of NR
simulations. As the field continues to evolve toward GPU-dominated
systems, the methods and tools presented here will play a pivotal role in
advancing gravitational-wave astrophysics and multi-messenger astronomy.


\section*{Acknowledgments}
ST gratefully acknowledges support from NASA award ATP-80NSSC22K1898 and
support from the University of Idaho P3-R1 Initiative.
LRW gratefully acknowledges support from NASA award LPS-80NSSC24K0360.
TA acknowledges support from NSF grants OAC-2229652 and AST-2108269, and from
the University of Wisconsin-Milwaukee.
ZBE's work was
supported by NSF grants OAC-2004311, OAC-2411068, AST-2108072,
PHY-2110352/2508377, and PHY-2409654, as well as NASA
ATP-80NSSC22K1898 and TCAN-80NSSC24K0100. This research made use of
Idaho National Laboratory's High Performance Computing systems located at
the Collaborative Computing Center and supported by the Office of Nuclear
Energy of the U.S. Department of Energy and the Nuclear Science User
Facilities under Contract No. DE-AC07-05ID14517. Finally, this work
benefited from the extensive use of the open-source packages
NumPy~\cite{NumPy}, SciPy~\cite{SciPy}, SymPy~\cite{SymPy}, and
Matplotlib~\cite{Matplotlib}.

\section*{Code Availability}
The latest version of \nellg is available at: \\
\url{https://doi.org/10.5281/zenodo.15115503}.
\newpage
\section*{Glossary}
\begin{table}[h!]
  \centering
  \begin{tabular}{ll}
    \toprule
    \textbf{Acronym} & \textbf{Definition} \\
    \midrule
    3G & Third Generation \\
    AI & Arithmetic Intensity \\
    API & Application Programming Interface \\
    AMReX & Adaptive Mesh Refinement for Exascale \\
    BC & Kernel that applies boundary conditions \\
    BBH & Binary Black Hole \\
    BSSN & Baumgarte-Shapiro-Shibata-Nakamura formulation \\
    CFL & Courant-Friedrichs-Lewy \\
    CPU & Central Processing Unit \\
    CSE & Common Subexpression Elimination \\
    DFMA & Double-Precision Fused Multiply-Add \\
    DADD & Double-Precision Add \\
    DRAM & Dynamic Random Access Memory \\
    ETK & Einstein Toolkit \\
    GFLOP/s & Gigaflops per Second \\
    GPU & Graphics Processing Unit \\
    GRMHD & General Relativistic Magnetohydrodynamics \\
    GW & Gravitational Wave \\
    H & Kernel that computes Hamiltonian constraints \\
    HPC & High-Performance Computing \\
    ID & Initial Data \\
    LISA & Laser Interferometer Space Antenna \\
    NR & Numerical Relativity \\
    NRPy & Numerical Relativity in Python \\
    RHS & Kernel to compute the Right-Hand-Side \\
    RK & Kernel to compute Runge-Kutta sub-step \\
    SIMD & Single Instruction, Multiple Data \\
    SIMT & Single Instruction, Multiple Threads \\
    SM & Streaming Multiprocessor \\
    TDP & Thermal Design Power \\
    \bottomrule
  \end{tabular}
  \label{tab:acronyms}
  \caption{Glossary of acronyms used throughout the paper.}
\end{table}
\newpage
\printbibliography
\end{document}